\documentclass[12pt,preprint]{aastex}
\usepackage[onecolumn,numberedappendix]{emulateapj5}

\usepackage{epsfig}
\usepackage{multirow}
\usepackage{graphics}
\usepackage{amsmath, amsthm, amssymb}
\usepackage{rotating}
\usepackage{lscape}
\tightenlines
\slugcomment{Accepted to the Astrophysical Journal on 31st Oct. 2013}

\newcommand \lsim{\mathrel{\rlap{\lower4pt\hbox{\hskip1pt$\sim$}}
    \raise1pt\hbox{$<$}}}
\newcommand \gsim{\mathrel{\rlap{\lower4pt\hbox{\hskip1pt$\sim$}}
    \raise1pt\hbox{$>$}}}

\newcommand     \kms    {\,{\rm km~s}^{-1}}

\newcommand{\beq}{\begin{equation}}
\newcommand{\eeq}{\end{equation}}
\newcommand{\beqa}{\begin{eqnarray}}
\newcommand{\eeqa}{\end{eqnarray}}

\newcommand{\phig}{\phi_{\rm geom}}

\newcommand{\phipb}{\phi_{{\bar P}}}
\newcommand{\phipc}{\phi_{P,c}}

\newcommand{\alvcl}{\alpha_{\rm vir,cl}}

\newcommand{\psc}{P_{s,c}}

\newcommand{\nnhp} {${\rm N_2H^+}$}
\newcommand{\nndp} {${\rm N_2D^+}$}

\newlength{\figwidth}
\countdef\decade=200
\advance\decade by \year
\countdef\hours=201
\advance\hours by \time
\divide\hours by 60
\countdef\mins=202
\advance\mins by \hours
\multiply\mins by 60
\multiply\hours by 100
\countdef\miltime=203
\advance\miltime by \hours
\advance\miltime by \time
\advance\miltime by -\mins

\begin{document}

\title{The Dynamics of Massive Starless Cores with ALMA}

\author{Jonathan C. Tan}
\affil{Departments of Astronomy \& Physics, University of Florida, Gainesville, FL 32611, USA}
\author{Shuo Kong, Michael J. Butler}
\affil{Department of Astronomy, University of Florida, Gainesville, FL 32611, USA}
\author{Paola Caselli}
\affil{School of Physics \& Astronomy, The University of Leeds, Leeds, LS2 9JT, UK}
\author{Francesco Fontani}
\affil{INAF - Osservatorio Astrofisico di Arcetri, Largo Enrico Fermi 5, I - 50125 Firenze, Italy}

\begin{abstract}
How do stars that are more massive than the Sun form, and thus how is
the stellar initial mass function (IMF) established? Such intermediate-
and high-mass stars may be born from relatively massive pre-stellar
gas cores, which are more massive than the thermal Jeans mass. The
Turbulent Core Accretion model invokes such cores as being in
approximate virial equilibrium and in approximate pressure equilibrium
with their surrounding clump medium. Their internal pressure is
provided by a combination of turbulence and magnetic fields.
Alternatively, the Competitive Accretion model requires strongly
sub-virial initial conditions that then lead to extensive
fragmentation to the thermal Jeans scale, with intermediate- and
high-mass stars later forming by competitive Bondi-Hoyle accretion. To
test these models, we have identified four prime examples of massive
($\sim100\:M_\odot$) clumps from mid-infrared extinction mapping
of infrared dark clouds (IRDCs). Fontani et al. found high deuteration
fractions of \nnhp in these objects,
which are consistent with them being starless.
Here we present ALMA observations of these four clumps that probe
the \nndp(3-2) line at 2.3\arcsec\ resolution. We find six \nndp cores
and determine their dynamical state. Their observed velocity
dispersions and sizes are broadly consistent with the predictions of
the Turbulent Core model of self-gravitating, magnetized (with
Alfv\'en Mach number $m_A\sim1$) and virialized cores that are bounded
by the high pressures of their surrounding clumps. However, in the
most massive cores, with masses up to $\sim60\:M_\odot$, our results
suggest that moderately enhanced magnetic fields (so that
$m_A\simeq0.3$) may be needed for the structures to be in virial and
pressure equilibrium. Magnetically regulated core formation may thus
be important in controlling the formation of massive cores, inhibiting
their fragmentation, and thus helping to establish the stellar IMF.
\end{abstract}


\section{Introduction}\label{S:intro}

The two main theories for massive (and intermediate-mass) star
formation, Core Accretion (e.g. McLaughlin \& Pudritz 1996; McKee \&
Tan 2003, hereafter MT03) and Competitive Accretion (e.g. Bonnell et
al. 2001; Wang et al. 2010), invoke very different initial conditions
for the gas about to collapse to form a high-mass star. The Turbulent
Core Accretion model of MT03 starts with massive
near-virial-equilibrium starless {\it cores} that will collapse to
form individual stars or close binaries. These cores can be considered
to be scaled-up versions of the much more common low-mass cores known
to form low-mass stars (Shu et al. 1987). The main differences from
low-mass cores are that the pressure support must be provided by
nonthermal forms (turbulence and/or magnetic fields) and the typical
environments where massive cores and stars form, self-gravitating
massive {\it clumps} (i.e., proto-star-clusters), are at much higher
pressure. Competitive Accretion also involves fragmentation of massive
gas clumps, but now into protostellar seeds with initial masses only
of order the thermal Jeans mass --- typically much less than a solar
mass under these high pressure conditions. Those protostellar seeds
that happen to be in high density regions then later accrete
previously unbound gas to eventually become intermediate-mass and
high-mass stars. The numerical simulations of Bonnell et al. (2001),
from which the competitive accretion theory was developed, involved
global, near free-fall collapse of a gas clump, which either started
at or evolved to a very sub-virial state. Krumholz et al. (2005)
showed such a sub-virial state was needed for the competitive
accretion rate to be large enough to be relevant for massive star
formation, i.e. allowing formation within timescales $\lesssim
1$~Myr. Thus to distinguish between these theoretical models we need
to find and measure the dynamical state of intermediate-mass and
high-mass starless cores: how close are they to virial equilibrium?

Answering this question is challenging because massive starless cores
are rare, typically far away, small in angular size, usually
surrounded by large quantities of other molecular gas, and expected to
suffer depletion of many molecular species, such as CO and CS, that
are often used to measure the mass and kinematics of molecular
clouds. Furthermore, it is difficult to infer the degree of support
a core receives from large scale magnetic fields.

\subsection{Target selection of intermediate-mass and high-mass starless cores}

To find massive starless cores we started by studying infrared dark
clouds (IRDCs) (e.g. P\'erault et al. 1996; Egan et al. 1998) using
{\it Spitzer} IRAC 8$\rm \mu m$ GLIMPSE survey (Benjamin et al. 2003)
images. We developed a mid-infrared extinction (MIREX) mapping
technique to derive mass surface densities, $\Sigma$, of clouds,
probing up to $\Sigma\simeq0.5\:{\rm g\:cm^{-2}}$ (i.e., $N_{\rm H}=
2.1\times 10^{23}\:{\rm cm^{-2}}$) and on scales down to $2\arcsec$
(Butler \& Tan 2009, hereafter BT09; Butler \& Tan 2012, hereafter
BT12). The method depends on the dust opacity but not the dust
temperature.
Figure~\ref{fig:maps}a shows $\Sigma$ maps of 4 core/clumps
(C1, F1, F2, G2) selected from the larger sample of 42 studied by
BT12. The properties of these objects are also listed in
Table~\ref{tab:clumps}. Selection of these particular sources was
guided by them having large values of $\Sigma$, still being dark at 24
and 70~$\rm \mu m$ ({\it Spitzer} MIPSGAL (Carey et al. 2009) images were
analyzed), and being in relatively quiescent environments with respect
to other star formation activity (MIR sources are absent in a $\sim
20\arcsec$ diameter aperture around the $\Sigma$ peak).

\begin{deluxetable}{cccccc}
\tabletypesize{\footnotesize}
\tablecolumns{6}
\tablewidth{0pt}
\tablecaption{IRDC Core/Clumps Observing Targets}
\tablehead{
\colhead{Core name\tablenotemark{a}} & \colhead{RA(J2000)} & \colhead{Dec(J2000)} & \colhead{$V_{\rm LSR}$\tablenotemark{b}} & \colhead{$d$\tablenotemark{c}} & \colhead{$D_{\rm frac}$\tablenotemark{b}}\\
\colhead{}    & \colhead{(h m s)} & \colhead{($^{\circ}\:\:\arcmin\:\:\arcsec$)} & \colhead{($\rm km\:s^{-1}$)} & \colhead{(kpc)} & \colhead{}}
\startdata
G028-C1 & 18:42:46.9 & -04:04:08 & +78.3 & 5.0 & 0.38\\
G034-F1\tablenotemark{d} & 18:53:16.5 & +01:26:10 & +57.7 & 3.7\tablenotemark{e} & 0.40\\
G034-F2\tablenotemark{d} & 18:53:19.1 & +01:26:53 & +57.7 & 3.7\tablenotemark{e} & 0.43\\
G034-G2 & 18:56:50.0 & +01:23:08 & +43.6 & 2.9 & 0.70\\
\enddata
\tablenotetext{a}{From BT12}
\tablenotetext{b}{From Fontani et al. (2011)}
\tablenotetext{c}{Kinematic distance from Simon et al. (2006)}
\tablenotetext{d}{The labeling of F1 and F2 was mistakenly interchanged in Fontani et al. (2011).}
\tablenotetext{e}{Kurayama et al. (2011) report a parallax distance of 1.56~kpc for IRDC F, although the reliability of this has been questioned by Foster et al. (2012).}
\label{tab:clumps}
\end{deluxetable}

\subsection{CO Freeze-out and Deuterium Fractionation} 

In these cold ($T\sim 15$~K; Pillai et al. 2006), high density ($n_{\rm
  H}\gtrsim 10^5\:{\rm cm^{-3}}$; BT12) environments, freeze-out of
molecules such as CO on to dust grains is expected, as has been
observed in low-mass starless cores (e.g. Caselli et
al. 1999). Hernandez et al. (2011) found widespread CO depletion in
the filamentary IRDC H of the BT12 sample. It has also been reported
in other similar clouds by, e.g. Zhang et al. (2009) and Fontani et
al. (2012). Molecules suffering freeze-out are poor tracers of
starless core kinematics. However, as CO is depleted from the gas
phase, the level of deuteration of certain species is expected to
increase (Dalgarno \& Lepp 1984). This is because CO reacts with and
thus destroys $\rm H_2D^+$ (that has formed by reaction of $\rm H_3^+$
with HD), so when CO is depleted the concentration of $\rm H_2D^+$
builds up and so deuteration of other species still present in the
gas phase, such as $\rm N_2H^+$, becomes enhanced.

In fact the 4 IRDC clumps we selected were found to have the highest
levels of deuteration, $D_{\rm frac} \equiv N_{\rm N_2D^+}/N_{\rm
  N_2H^+}$ among the 10 massive starless clouds studied by Fontani et
al. (2011), with $D_{\rm frac}$ = 0.38, 0.43, 0.40, 0.70 for C1, F1,
F2, G2, respectively. This compares with a mean $D_{\rm frac}=0.12$
for the other 6 sources in the sample, which were not selected from
MIR extinction. This indicates that the MIREX-based selection method
is efficient at finding regions that are cold and dense, and not yet
forming stars. Note that the protostellar sources in the sample of
Fontani et al. tend to have significantly lower values of $D_{\rm
  frac}\simeq 0.04$, which is expected due to the warming of the
infall envelope by the protostar.


Deuterated molecules have been shown to be the best kinematic probes
of the coldest and densest conditions associated with low-mass star
formation (e.g. Caselli et al. 2002; Crapsi et al. 2007). We will
apply the same methods in the more extreme environments of IRDCs,
where there tends to be a larger and more pressurized mass of cold,
dense gas, compared to low-mass star-forming regions such as Taurus.
We will thus use \nndp to both identify pre-stellar cores ---
efficiently distinguishing them from their surrounding clump material
--- and then as their kinematic tracer.

\subsection{Predictions of the Turbulent Core Accretion Model of Virialized Cores}\label{S:theory}

MT03 modeled clumps and massive pre-stellar cores as singular
polytropic virialized spheres in hydrostatic equilibrium, with power
law density, $\rho \propto r^{-k_{\rho}}$, and pressure, $P\propto
r^{-k_{P}}$, distributions with fiducial value $k_{\rm \rho}= 1.5$
(for both clumps and cores), implying $k_{P}=2(k_{\rho}-1)=1.0$. Such
values are consistent with those derived from MIREX mapping by BT12,
who found $k_{\rm \rho,cl}\simeq 1.1$ (for clumps) and $k_{\rm
  \rho,c}\simeq 1.6$ (for cores). MT03 assumed cores have a boundary
with surface pressure, $P_{s,c}$, that is in approximate equilibrium
with the pressure of the immediately surrounding environment of the
star-forming clump. This pressure was assumed to be related to the
mean pressure in the clump, $\bar{P}_{\rm cl}$, via $\psc = \phipc
\bar{P}_{\rm cl}$, where $\phipc$ is a constant of order unity. The
value of $\phipc$ depends on the location of the core within the
clump: MT03 assumed cores were typically at radial locations of
$0.3R_{\rm cl}$ so that $\phipc\simeq 2$ (for $k_{\rm
  \rho,cl}=1.5$). In this paper we will define the clump properties
from the material that is more localized around each core: we set
$R_{\rm cl}=2 R_c$ and evaluate its mass surface density, $\Sigma_{\rm
  cl}$, as the average value (observed via the MIREX maps) in the
annulus from $R_c$ to $2R_c$. In this case, $\phipc =
(1-k_P/3)(R_c/R_{\rm cl})^{-k_P} \rightarrow 4/3$ in the fiducial case
of $k_\rho=1.5, k_P=1$ (and 1.07 in the case of $k_\rho=1.1$, $k_P=0.2$).

MT03 assumed the mean pressure in the clump was set by its
self-gravitating weight, i.e. $\bar{P}_{\rm cl}\equiv \phipb
G\Sigma_{\rm cl}^2$, where $\phipb = (3\pi/20)f_g\phig \phi_B \alvcl$,
where $f_g$ is the gas mass fraction of the clump, $\phig$ is a
numerical factor of order unity that accounts for the effect of
nonspherical geometry, $\phi_B$
accounts for the effect of magnetic fields and $\alvcl\equiv 5\langle
\sigma_{\rm cl}^2 \rangle R_{\rm cl}/(GM_{\rm cl})$ is the virial
parameter of the clump. MT03 adopted a fiducial value of $f_g=2/3$
(with the remainder composed of a nascent stellar cluster), but for
the IRDC clumps we consider here, we will assume $f_g=1$. Aspect
ratios of up to 2:1 (i.e., eccentricity of a projected elliptical
shape $e\leq 0.87$) lead to $\phig \lesssim 1.1$, so such elongations
are a relatively minor effect and, like MT03, we assume $\phig=1$. For
magnetic field strengths such that the Alfv\'en Mach number $m_A
\equiv \sqrt{3} \sigma/v_A =1$, where $v_A = B/\sqrt{4\pi\rho}$ is the
Alfv\'en speed, MT03 showed that $\phi_B=1.3 + 1.5m_A^{-2}\rightarrow
2.8$. Below, for our fiducial estimates, we will consider the
possibility of a range $0.5<m_A<2$, corresponding to
$7.3>\phi_B>1.7$. Finally, we assume that the clump is virialized with
$\alvcl=1$. With these values $\phipb=1.32$ (c.f. 0.88 in MT03).

The radius of a core in virial equilibrium, including pressure equilibrium with its surroundings, is (MT03)
\begin{eqnarray}
R_{\rm c,vir} & = & 0.071 \left(\frac{A}{k_P^2\phipc \phipb}\right)^{1/4} \left(\frac{M_c}{60\:M_\odot}\right)^{1/2}\left(\frac{\Sigma_{\rm cl}}{1\:{\rm g\:cm^{-2}}}\right)^{-1/2}\:{\rm pc}\\
R_{\rm c,vir} & \rightarrow & 0.0574 \left(\frac{M_c}{60\:M_\odot}\right)^{1/2}\left(\frac{\Sigma_{\rm cl}}{1\:{\rm g\:cm^{-2}}}\right)^{-1/2}\:{\rm pc},
\label{eq:rcore}
\end{eqnarray}
where $A=(3-k_\rho)(k_\rho-1)f_g\rightarrow 3/4$. The mass average
velocity dispersion of a virialized core is related to that at the
surface via $\sigma_{\rm c,vir} = [2(3-k_\rho)/(8-3k_\rho)]\sigma_{\rm
  c,s} \rightarrow (6/7)\sigma_{\rm c,s}$. Note that $\sigma_{\rm c} =
c_{\rm c} / \phi_B^{1/2}$, where $c_c\equiv (P_c/\rho_c)^{1/2}$ is the effective sound speed in
the core, so that
\begin{eqnarray}
\sigma_{\rm c,vir} & = & 1.91 \frac{2(3-k_\rho)}{8-3k_\rho} \left(\frac{\phipc\phipb}{A k_P^2\phi_B^4}\right)^{1/8}
\left(\frac{M_c}{60 M_\odot}\right)^{1/4}
\left(\frac{\Sigma_{\rm cl}}{1\:{\rm g\:cm^{-2}}}\right)^{1/4}\:{\rm km\:s^{-1}}\\
\sigma_{\rm c,vir} & \rightarrow & 1.09
\left(\frac{M_c}{60 M_\odot}\right)^{1/4}
\left(\frac{\Sigma_{\rm cl}}{1\:{\rm g\:cm^{-2}}}\right)^{1/4}\:{\rm km\:s^{-1}}.\label{eq:s2}
\end{eqnarray}
The above expressions can also be combined to give a
velocity dispersion (or FWHM line width, $\Delta v_{\rm vir} = (8 {\rm ln}
2)^{1/2}\sigma_{\rm vir} = 2.355 \sigma_{\rm vir}$) versus size relation:
\begin{eqnarray}
\sigma_{\rm c,vir} & = & 2.27 \frac{2(3-k_\rho)}{8-3k_\rho} \left(\frac{\phipc\phipb}{A \phi_B^2}\right)^{1/4} \left(\frac{\Sigma_{\rm cl}}{1\:{\rm g\:cm^{-2}}}\right)^{1/2}  \left(\frac{R_{\rm c,vir}}{0.1\:{\rm pc}}\right)^{1/2}\:{\rm km\:s^{-1}}\\
\sigma_{\rm c,vir} & \rightarrow & 1.44 \left(\frac{\Sigma_{\rm cl}}{1\:{\rm g\:cm^{-2}}}\right)^{1/2}  \left(\frac{R_{\rm c,vir}}{0.1\:{\rm pc}}\right)^{1/2}\:{\rm km\:s^{-1}}.
\label{eq:LWS}
\end{eqnarray}
We will now compare these predictions, in particular for $\sigma_{\rm
  c,vir}$ and $R_{\rm c,vir}$, with our sample of massive starless
cores.


\section{Observations}

We used ALMA in Cycle 0 compact configuration to observe $\rm
N_2D^+$(3-2) (231~GHz, Band 6, $\simeq$2.3\arcsec\ angular resolution,
$\sim 9\arcsec$ maximum detectable scale, 27\arcsec\ field of view,
0.08~$\rm km\:s^{-1}$ velocity resolution) with a single pointing for
each core, centered at the peak $\Sigma$ of the BT12 MIREX map, and
sharing one track amongst the 4 sources with a total integration time
of about 1.0 hour per source. Several other species, $\rm DCO^+$(3-2),
$\rm DCN$(3-2), $^{13}$CS(5-4) and SiO(5-4) were also searched for in
the same spectral setup. We achieved $1\sigma$ RMS noise levels of
about 11 and 7.0~mJy/beam per 0.08~km/s channel for $\rm N_2D^+$(3-2)
(231.32186~GHz) and DCO$^+$(3-2) (216.11258~GHz), respectively.

From the line-free regions of the four spectral
windows we measured the continuum flux with an average wavelength of
1.338~mm. The values of the $1\sigma$ RMS noise per beam in C1, F1, F2, G2 were
0.272, 0.213, 0.336, 0.174~mJy/beam, respectively.

\section{Results}

\subsection{Identification of starless cores via $\rm N_2D^+$ emission}\label{S:identification}

\begin{figure*}[!tb]
\begin{center}
\includegraphics[width=7.2in,angle=0]{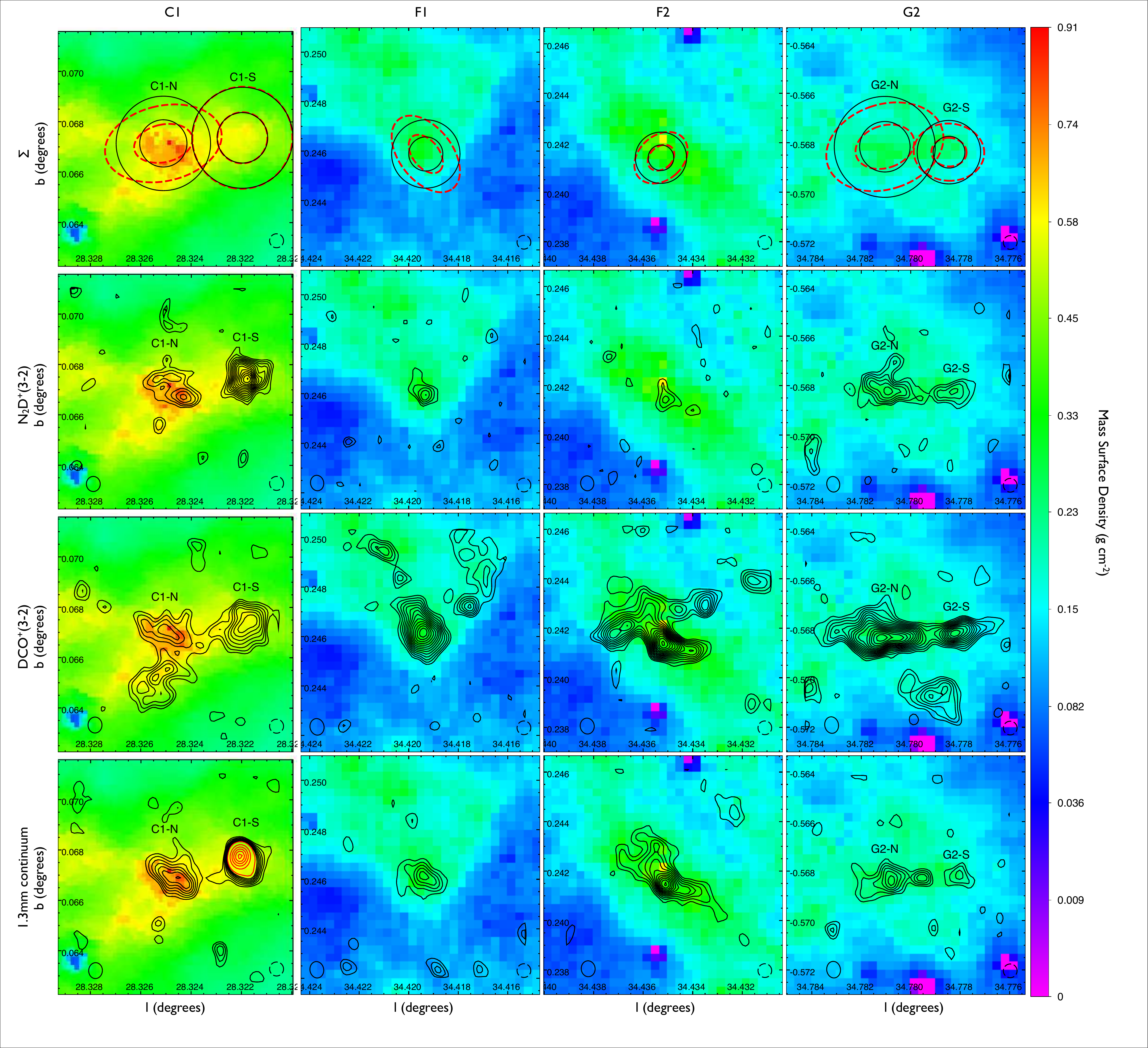} 
\end{center}
\caption{\footnotesize
The four massive starless core/clumps observed by ALMA: columns from
left to right show C1, F1, F2, G2. The background colors in all rows
show MIREX $\Sigma$ maps in ${\rm g\:cm^{-2}}$ (Butler, Tan \&
Kainulainen 2013 for C1; BT12 for F1, F2 and G2). The {\it
  Spitzer}-IRAC 2\arcsec\ beam is shown in the lower right of the
panels.  {\it (a) First row:}
Analytic regions used to define the cores and their surrounding clump
environments.  The inner ellipses and equivalent area circles show the
deconvolved extent of the \nndp cores identified in row (b), while the
outer ellipses and circles have a radius twice as large and define
the annuli used to estimate the surrounding clump envelope.
{\it (b) Second row:} ALMA Cycle 0 observations of \nndp(3-2)
integrated intensity, contours shown from $2, 3, 4 ... \sigma$. Six
cores, C1-N, C1-S, F1, F2, G2-N \& G2-S, are defined by their
3$\sigma$ contours in position-velocity space.
The ALMA beam is shown in the lower left of each panel.
Note that not all high $\Sigma$ regions show strong \nndp emission,
but \nndp cores do correlate well with structures seen in the MIREX
maps.  {\it (c) Third row:} ALMA observations of DCO$^+$(3-2)
integrated intensity, contours shown from $2, 3, 4 ... \sigma$, which
is generally more widespread than \nndp. {\it (d) Fourth row:} ALMA
observations of the 1.34~mm dust continuum emission, with contours
from $2, 3, 4 ... 8 \sigma$ (black) then (for C1) 10, 20, 30 ... 60$\sigma$
(red).}
\label{fig:maps}
\end{figure*}

The integrated intensity maps of $\rm N_2D^+$(3-2) and DCO$^+$(3-2) of
the clumps C1, F1, F2, G2 are shown in Fig.~\ref{fig:maps}b \& c as
contour plots overlaid on top of the MIREX $\Sigma$ maps. Strong
detections of both lines are seen in all cases. $\rm DCN$(3-2),
$^{13}$CS(5-4) and SiO(5-4) were not detected, even from stacked
rest-frame spectra towards the $\rm N_2D^+$ cores. Continuum emission
from the line-free regions of the four spectral windows, with mean
wavelength of 1.338~mm, was detected and is shown in Fig.~1d.

The critical density of $\rm N_2D^+$(3-2) is $2.9\times 10^6\:{\rm
  cm^{-3}}$, while that of DCO$^+$(3-2) is $3.5\times 10^6\:{\rm
  cm^{-3}}$. However, the emission from DCO$^+$ tends to be more
widespread. We expect this is because DCO$^+$ (formed via $\rm CO +
H_2D^+$) is following the more extended distribution of CO, which is likely to
suffer significant depletion in the cold, high-density cores, where
$\rm N_2D^+$ is enhanced. There is generally good correspondence
between the morphology of the line emission and structures seen in the
MIREX maps. For example, the ``V''-shaped structure of DCO$^+$
emission in F1 is also seen in the MIREX map. Even quite weak
DCO$^+$ features, e.g. in G2, can show themselves via MIR extinction.

Some, but not all, localized MIREX column density structures reveal
themselves via $\rm N_2D^+$(3-2). Following the results of studies of
low-mass starless cores described in \S\ref{S:intro}, we define these
to be starless core candidates, finding 6 objects (C1-N, C1-S, F1, F2,
G2-N, G2-S) that have emission exceeding the 3$\sigma$ noise level
within $l-b-v$-space. Positional core boundaries are defined using the
projection of this 3$\sigma$ contour, after deconvolving with a
Gaussian equivalent to the ALMA synthesized beam. The positions,
elliptical eccentricities and position angles, and equivalent area
radial sizes, $R_c$, are listed in Table~\ref{tab:2}. The diameters of
the cores are all larger than the angular resolution of the
observations. Core C1-N appears to exhibit some substructure in its
$\rm N_2D^+$(3-2) emission, whereas the other sources appear to be
relatively monolithic.

The morphology of the continuum emission also generally matches that
seen in the molecular line emission, especially that traced by $\rm
N_2D^+$(3-2).

\subsection{Core kinematics and velocity dispersion}

\begin{figure*}[!tb]
\begin{center}
\includegraphics[width=7.2in,angle=0]{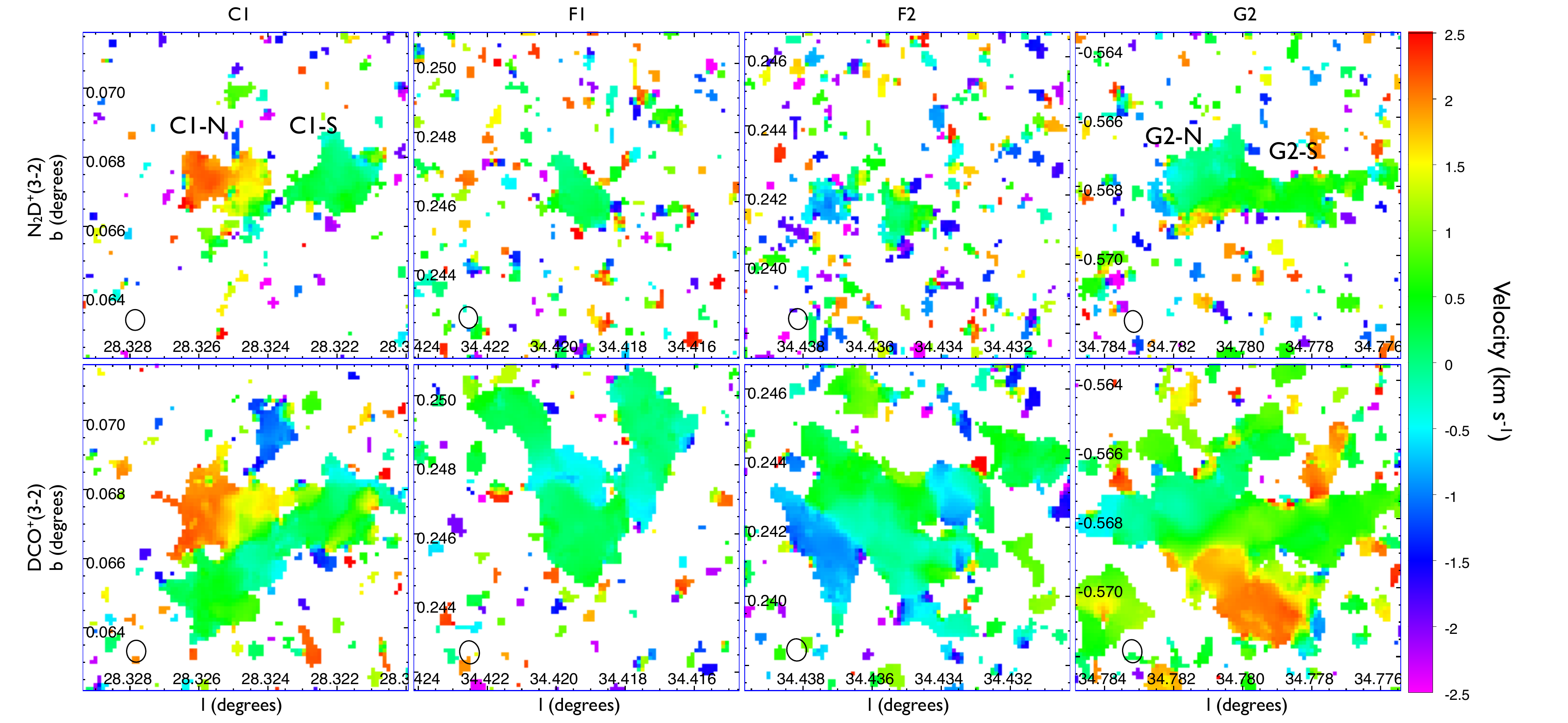}
\end{center}
\caption{
First moment (mean velocity) maps of the four core/clumps: columns
from left to right show C1, F1, F2, G2. {\it (a) First row:} First
moment map of \nndp(3-2) emission, integrated over a 5 km/s velocity
range centered on the mean \nndp core velocities of C1-S, F1, F2 and
G2-N. The ALMA beam is shown in the lower left of each panel. Only
regions that have some contribution from cells with intensities
$\geq3\sigma$ are displayed in color. {\it (b) Second
  row:} First moment map of DCO$^+$(3-2) emission in the same velocity
reference frames as in row (a).
\label{fig:mom1}}
\end{figure*}

In Fig.~\ref{fig:mom1} we show the first moment maps of the 4
core/clumps as traced by $\rm N_2D^+$(3-2) and DCO$^+$(3-2). We only
include the contribution from cells in the position-velocity cube that
have a signal more than $3\sigma$ above the noise in each 0.08~km/s
channel. We integrate over a velocity range of $\pm 2.5\:{\rm
  km\:s^{-1}}$ centered on the mean velocity of the $\rm N_2D^+$ cores
C1-S, F1, F2, and G2-S (evaluated below). In the C1 region, cores C1-N
and S show a velocity offset of 1.8$\:{\rm km\:s^{-1}}$, but with some
$\rm N_2D^+$ emission at intermediate velocities (we checked that all
significant emission associated with C1-N is captured by the velocity
interval used in Fig.~\ref{fig:mom1}). In G2, the cores are within
0.4$\:{\rm km\:s^{-1}}$ of each other. On the larger scales of the
DCO$^+$(3-2) emission the first moment maps generally reveal quite
disordered structure, with multiple discrete features spanning a
velocity range of a few ${\rm km\:s^{-1}}$. The exception is F1, which
shows quite coherent velocity structure across its ``V''-shaped
morphology.

\begin{figure*}[!tb]
\begin{center}
\includegraphics[width=7.2in,angle=0]{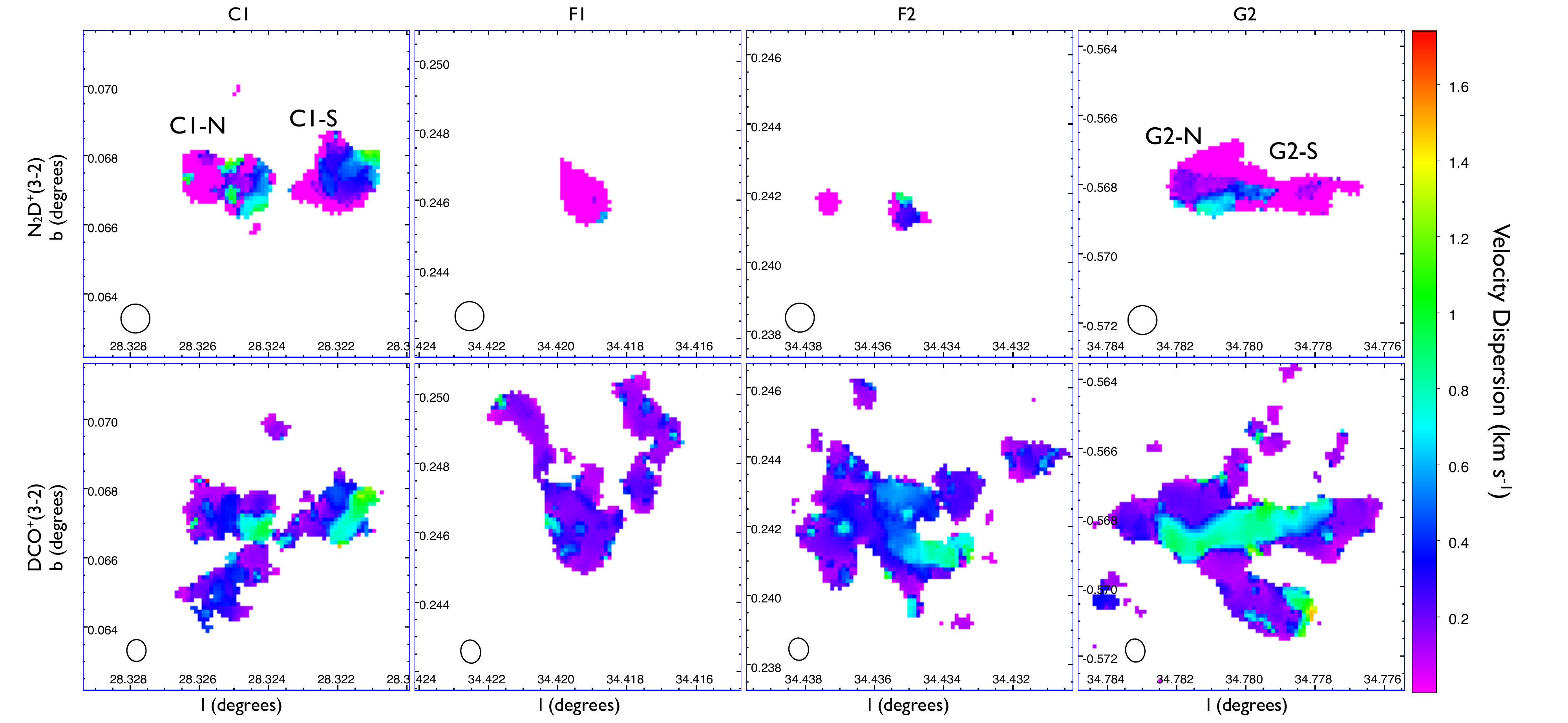}
\end{center}
\caption{
Second moment (velocity dispersion $\sigma_{\rm N_2D^+,obs}$ and
$\sigma_{\rm DCO^+,obs}$) maps of the four core/clumps: columns from
left to right show C1, F1, F2, G2. {\it (a) First row:} Second moment
map of \nndp(3-2) emission, smoothed with a 3\arcsec\ beam, integrated
over a 5 km/s velocity range centered on the mean \nndp core
velocities of C1-S, F1, F2 and G2-S. Only regions meeting signal to
noise threshold requirements to measure velocity dispersion (see text)
are displayed in color. {\it (b) Second row:} Second moment map of
DCO$^+$(3-2) emission (unsmoothed) in the same velocity reference frames as in row
(a).
\label{fig:mom2}}
\end{figure*}

In Fig.~\ref{fig:mom2} we show the second moment (velocity dispersion)
maps of the four core/clumps. To evaluate these, only regions that
have at least one cell in the velocity range used for the first moment
map with intensity $\geq 5\sigma$ are considered. Then the velocity
dispersion is evaluated from those cells that have signal $\geq
3\sigma$. This truncation will tend to lead to an underestimation of
the true dispersion, especially for regions with weak signal (below,
when we evaluate the velocity dispersion of cores with defined areas,
we will do so by fitting directly to the total spectrum). For the $\rm
N_2D^+(3-2)$ emission we also subtract off in quadrature the
contribution to line broadening, 0.242$\kms$, from the main group of
hyperfine lines (over a velocity range of $\pm1\kms$ about the peak
line). This assumes the lines are optically thin, which is consistent
with our analysis of core spectra (below). This hyperfine broadening
can be larger than the observed velocity dispersion in regions with
relatively low signal to noise ratios, and in this case we set the map
at this location to have an unmeasured value. In order to have sufficient
signal to noise to see extended structure in the velocity dispersion
maps of some of the cores, we smoothed the $\rm N_2D^+$ map with a
3\arcsec\ beam.

Fig.~\ref{fig:mom2} reveals velocity dispersions that are typically a
fraction of a $\kms$. Low velocity dispersion halos around some
structures are likely caused by the signal to noise ratio threshold,
described above, that is used to select regions in position-velocity
space for analysis. Nevertheless, real structure can be seen in the
maps, especially those of DCO$^+$. Structures showing larger velocity
dispersion can be real, e.g. as the result of local virialized motions
in a core or clump, or can result from overlapping independent
components. A more detailed comparison of the observed velocity
structures with the results of numerical simulations of molecular
clouds will be carried out in a future paper.

\begin{figure*}[!tb]
\begin{center}
\includegraphics[width=3in,angle=0]{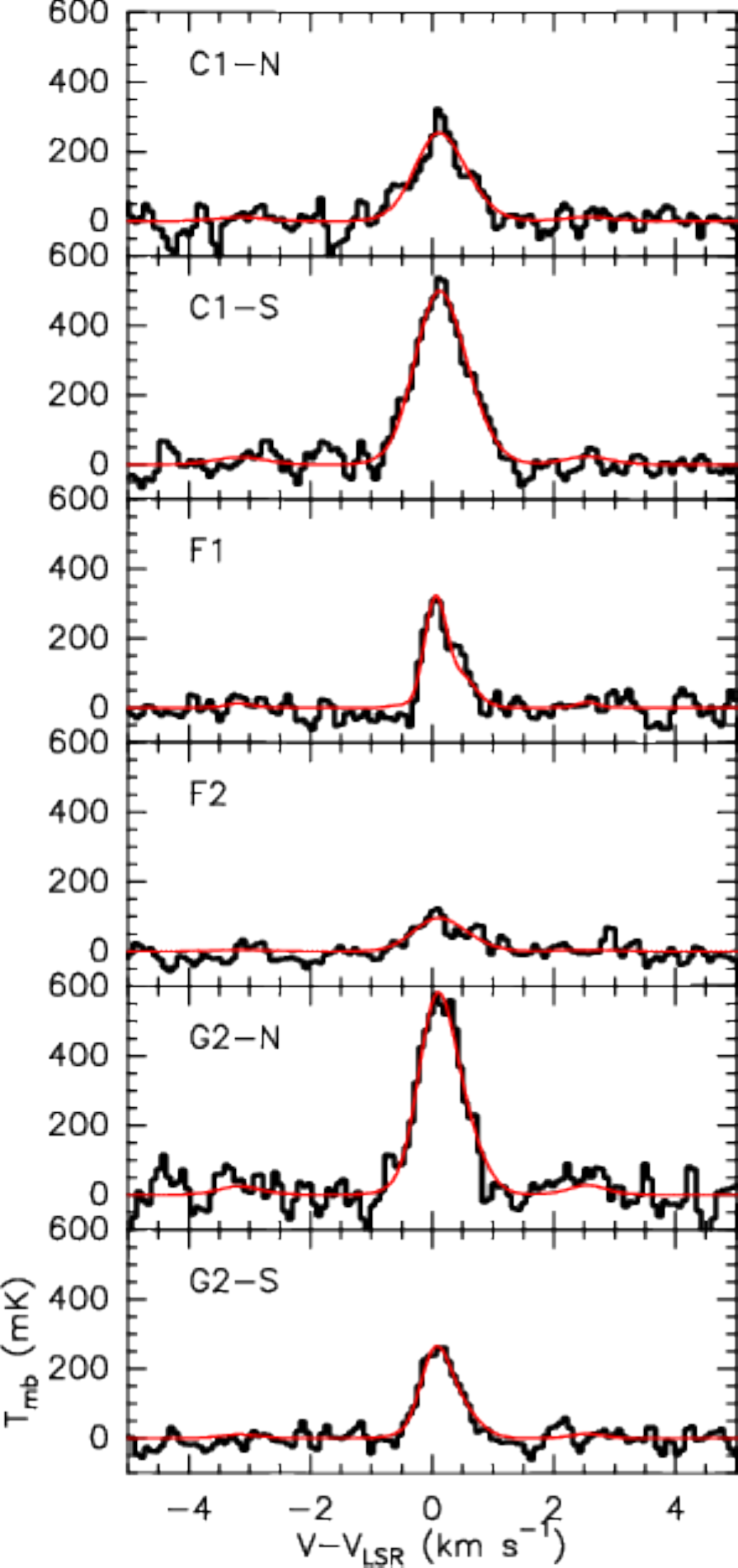} 
\end{center}
\caption{
\nndp(3-2) spectra ($V_{\rm LSR}$-frame) of the six identified cores: C1-N, C1-S, F1, F2, G2-N \& G2-S, shown at the maximum resolution of 0.08~${\rm km\:s^{-1}}$. The red continuous line shows the best fit model profiles.
}
\label{fig:spectra}
\end{figure*}

We now focus on the velocity dispersion of the identified $\rm N_2D^+$
cores.  The integrated $\rm N_2D^+$(3-2) spectra of the cores
identified in \S\ref{S:identification} are shown in
Fig.~\ref{fig:spectra}. We fit model spectra, which account for the
full blended set of 47 hyperfine components, to these data to derive
the centroid velocity, $V_{\rm LSR,N_2D^+}$, and the observed 1D
velocity dispersion, $\sigma_{\rm N_2D^+,obs}$, also listed in
Table~\ref{tab:2}. This modeling allows for the possibility of
optically thick parts of the hyperfine line complex, but we find all
spectra can be well-modeled assuming optically thin line emission.
Note, these values for the cores can appear larger than those
displayed in Fig.~\ref{fig:mom2}, since the former includes the
contribution from large-scale gradients in centroid velocity across
the core. We assume a gas temperature of $T=10\pm3$~K (see
\S~\ref{S:dustmass}) to remove the thermal component of the line to
thus assess the nonthermal component of the velocity dispersion. We
then sum this in quadrature with the sound speed of the molecular
cloud, assuming the same temperature value of $10\pm3$~K and a mean
particle mass of $\mu=2.33 m_p$, to derive the total core velocity
dispersion as derived from $\rm N_2D^+$(3-2) emission, $\sigma_{\rm
  N_2D^+}$.

We do not use DCO$^+$ to measure core kinematics since it is expected
to be somewhat under abundant in CO-depleted regions and thus likely
to preferentially trace the extended core envelopes. Nevertheless we
present in Table~\ref{tab:2} the values of $V_{\rm LSR,DCO^+}$ and
$\sigma_{\rm DCO^+,obs}$, integrated over the same region of the $\rm
N_2D^+$-defined core. The centroid velocities of $\rm N_2D^+$(3-2) and
DCO$^+$(3-2) are very similar (always within $0.2\:{\rm km\:s^{-1}}$),
while the observed velocity dispersion of DCO$^+$ tends to be larger
by up to $\sim 35\%$, which is probably related to this species
tracing a larger region around the core.


\subsection{Core and clump mass surface density, mass and density estimates}

\subsubsection{Estimates from MIREX maps}

We use the BT12 MIREX maps to estimate the mass surface density of the
clump material, $\Sigma_{\rm cl}$, immediately surrounding the
identified cores, from annuli extending from $R_c$ to $2R_c$.  This
choice of size of the annulus follows the method of BT12 (see also
\S\ref{S:theory}). We have investigated apertures based on the
elliptical shapes derived from the $\rm N_2D^+$ emission that define
the cores and based on equivalent area circles. The resulting
differences in the derived quantities are typically very minor,
$\lesssim 10\%$, so for simplicity we only report results based on
equivalent area circular apertures.

We find values of $\Sigma_{\rm cl}\simeq 0.2 - 0.3\:{\rm g\:cm^{-2}}$
(Table~\ref{tab:2}). We assume these estimates have a 30\% uncertainty
due to the choice of opacity per unit gas mass in the IRDC material
(BT12 assume $\kappa_{\rm 8\mu m}=7.5\:{\rm cm^2\:g^{-1}}$ based on
the moderately coagulated thin ice mantle model of Ossenkopf \&
Henning 1994). However, in very high column density regions, the MIREX
method of BT12 will begin to underestimate the true value of $\Sigma$,
when the background intensity that makes it through the cloud becomes
comparable to the instrumental noise in the image. BT12 refer to these
as ``saturated'' regions. Only F2 is formally classified as
``saturated'' by BT12, but this is based on a global definition within
the IRDC and does not allow for modest variations in the foreground
intensity across the cloud.


Butler, Tan \& Kainulainen (2013) (BTK) have examined archival {\it
  Spitzer}-IRAC data on IRDC C that probe to significantly smaller
instrumental noise levels than the {\it Spitzer}-GLIMPSE images
analyzed by BT12. We have evaluated $\Sigma_{\rm cl}$ from these BTK
maps and find values that are 54\% and 53\% larger for C1-S and C1-N,
respectively. 
We will perform our dynamical analysis
for the following cases: (1) uniformly using the BT12 maps for all 6 cores; (2)
replacing the BT12 maps with the BTK maps for cores C1-S and C1-N
(these entries are shown in square brackets in Table~\ref{tab:2}).

We then use the MIREX maps to evaluate the maximum mass surface
density of the core, $\Sigma_{\rm c,max}$, averaged inside $R_c$.  In
one sense this is an upper limit for the core properties because it
assumes all the line of sight material in the IRDC is associated with
the core. However, because of potential saturation in the MIREX maps,
our estimates of the total mass surface density through the IRDCs in
these regions could be underestimates. From the BT12 maps, we find
$\Sigma_{\rm c,max}\simeq 0.2 - 0.4\:{\rm g\:cm^{-2}}$, only slightly
higher than the values of $\Sigma_{\rm cl}$. This suggests that either
only a small fraction of the total column of material along the line
of sight is associated with the core (with the rest being part of the
surrounding clump) or that the mass surface densities of the cores are
underestimated because of saturation in the MIREX maps. To assess this
latter possibility, for the C1 cores we also evaluate $\Sigma_{\rm
  c,max}$ from the BTK map, finding values that are 65\% and 58\%
larger for C1-S and C1-N, respectively, than those derived from the
BT12 maps. In the BTK map, these cores have values of $\Sigma_{\rm
  c,max}$ that are both 16\% larger than $\Sigma_{\rm cl}$ (compared
to 8.5\% and 13\% for C1-S and C1-N, respectively, in the BT12
maps). As a result of the uncertainty in assessing what fraction of
$\Sigma_{\rm c,max}$ is associated with the cores, after first
deriving core properties based simply on $\Sigma_{\rm c,max}$, we will
also consider two further methods to estimate $\Sigma_{\rm c}$.

Utilizing $\Sigma_{\rm c,max}$, the ``maximum'' core mass is
derived as $M_{\rm c,max}=\Sigma_{\rm c,max}\pi R_c^2$. We assume 20\%
distance uncertainties (we have adopted the kinematic distance
estimates of Simon et al. 2006), which leads to 50\% total
uncertainties in $M_{\rm c,max}$ (summing errors in
quadrature). However, IRDC F has a reported astrometric distance of
1.56~kpc (Kurayama et al. 2011), only 42\% of the kinematic
distance. We thus perform a separate analysis for the F cores using
this closer distance (these entries are shown in rounded brackets in
Table~\ref{tab:2}). Note however that Foster et al. (2012) have called
into question the validity of the astrometric distance, since the data
quality on which it is based are relatively poor and the implied
streaming motions if it is correct are very large ($\sim 30\:{\rm
  km\:s^{-1}}$).

The derived maximum core masses, $M_{\rm c,max}$, are in the range
from $\sim 1$ to $\sim 60\:M_\odot$. C1-N \& S have similar masses at the
upper end of this range (using the BTK maps) and have the potential to
form massive (i.e. $>8M_\odot$) stars. Next, G2-N has about
9~$M_\odot$, followed by the F cores (assuming the kinematic distance)
and then G2-S. These may be the progenitors of intermediate-mass or
even low-mass stars. Note, however, that these star-forming
environments are at significantly higher mass surface densities, and
thus pressures, than most previously studied regions of low-mass star
formation, such as in the Taurus molecular cloud.

With $\Sigma_{\rm c,max}$ and $R_c$ we can also estimate the maximum
volume density assuming spherical geometry. Table~\ref{tab:2} lists
values of the maximum total H nuclei number density as $n_{\rm
  H,c,max}= 3 M_{\rm c,max}/(4\pi R_c^3\mu_{\rm H})$, which given the
above assumptions have 36\% uncertainties. These values range from a
few~$\times 10^5\:{\rm cm^{-3}}$ to few~$\times 10^6\:{\rm cm^{-3}}$.


We now attempt to account for the overlying clump material that is not
associated with the cores. First, we assume this material has the same
column density as the surrounding clump evaluated from $R_c$ to
$2R_c$. We expect that subtracting off this clump envelope leads to a
minimum estimate of the core mass, since if the local IRDC geometry is
sheet-like or filamentary this would lead to an overestimate of the
clump material that is along the line of sight to the core. We
evaluate $\Sigma_{\rm c,min}=\Sigma_{\rm c,tot}-\Sigma_{\rm cl}$ and
list its values in Table~\ref{tab:2}. We also then derive $M_{\rm
  c,min}$ and $n_{\rm H,c,min}$. This method of envelope subtraction
substantially reduces the estimated core mass by factors of about five
to ten. The most massive cores, C1-S \& N now have masses of about
$10\:M_\odot$ (using the BTK maps). By this method, the estimated
densities are now $\lesssim 10^5\:{\rm cm^{-3}}$.

\subsubsection{Estimates from 1.34~mm dust continuum emission}\label{S:dustmass}

Additionally, we utilize the observed 1.338~mm continuum emission to
estimate core properties. Our ALMA Cycle 0 compact configuration
observations are sensitive to angular scales from $\sim 2\arcsec$ to
9\arcsec, so will tend to pick out core rather than clump material.
The total mass surface density corresponding to a given specific
intensity of mm continuum emission is
\beq
\Sigma_{\rm mm} = 6.03\times
10^{-3} \left(\frac{S_\nu/\Omega}{{\rm MJy/sr}}\right) \left(\frac{\kappa_\nu}
{0.01\:{\rm cm^2\:g^{-1}}}\right)^{-1} \lambda_{1.338}^3 \left[{\rm exp}\left(1.075 T_{d,10}^{-1}
  \lambda_{1.338}^{-1}\right)-1\right]\:{\rm g\:cm^{-2}},
\label{eq:Sigmamm}
\eeq
where $\lambda_{1.338}=\lambda/1.338\:{\rm mm}$ and
$T_{d,10}=T_d/10\:{\rm K}$. At this frequency, our preferred choice of
$\kappa_\nu = 5.95\times 10^{-3}\:{\rm cm^2\:g^{-1}}$, based on the
moderately coagulated thin ice mantle dust model of OH94 and assuming
a total (gas plus dust)-to-refractory-component-dust-mass ratio of 141 (Draine 2011).
OH94 dust models that are more coagulated have $\kappa_\nu$ that is
about 23\% larger. Overall we adopt a 30\% uncertainty in $\kappa_\nu$.

An estimate of the dust temperature is also needed. From observations
of ammonia inversion transitions, Pillai et al. (2006) measured IRDC
temperatures $\sim 15$~K. However, we expect such observations tend to
probe the lower density envelopes around the IRDC cores, as the
inversion transitions of $\rm NH_3$ have critical densities of only
$\sim 10^4\:{\rm cm^{-3}}$. The dust temperature can also be
constrained from the FIR spectral energy distribution. For example,
fitting {\it Herschel} PACS and SPIRE data, Peretto et al. (2010)
found temperatures of $\sim 10-12$~K in the central region of an IRDC.

Here we use the fact that some IRDC cores are seen in absorption at
wavelengths as long as $\sim 100\:{\rm \mu m}$ to place constraints on
the dust temperature. This method has the advantage of allowing us to
derive constraints on scales of the {\it Herschel} angular resolution, i.e.
down to 6\arcsec\ for the 70$\rm \mu m$ band. Approximating the
cloud as a 1D slab that has a total optical depth $\tau_\nu$ at
frequency $\nu$ and integrating the radiative transfer equation yields
\beq I_{\nu,1} = I_{\nu,0} e^{-\tau_\nu} + \int_0^{\tau_\nu}
e^{-(\tau_\nu - \tau_\nu^\prime)}(j_\nu/\kappa_\nu) d \tau_\nu^\prime,
\eeq where $I_{\nu,1}$ is the transmitted intensity towards the
observer from the near side of the cloud, $I_{\nu,0}$ is the
background incident intensity on the far side of the cloud, $j_\nu$ is
the emissivity of the cloud material (i.e. its dust), and $\kappa_\nu$
is the cloud opacity. Approximating the dust in the IRDC as being in
thermal equilibrium and at a constant temperature so that
$j_\nu/\kappa_\nu\rightarrow B_\nu (T_d)\equiv (2h\nu^3/c^2)(e^{h\nu/kT_d}-1)^{-1}$, the Planck function, we
have \beq \frac{I_{\nu,1}}{I_{\nu,0}}= e^{-\tau_\nu} +
\frac{B_\nu(T_d)}{I_{\nu,0}} (1-e^{-\tau_\nu}). \eeq

The background intensity can be estimated empirically from the
observed emission in the Galactic plane. Li \& Draine (2001)
considered Galactic infrared emission in the MIRS region
($44^\circ\leq l \leq 44^\circ40^\prime, -0^\circ40^\prime\leq b \leq
0^\circ$). Integrating their model over the {\it Herschel}-PACS bands
at 70, 100, 160~$\rm \mu m$ gives intensities $I_\nu = 500, 1270,
1690\:{\rm MJy\:sr^{-1}}$ at mean wavelengths of 74.0, 103.6,
161.6~$\rm \mu m$, respectively. At $l=30^\circ,b=0^\circ$, Bernard et
al. (2010) estimated ``offset'' intensities of 688.0 and
1982.3~$\:{\rm MJy\:sr^{-1}}$ for the PACS 70 and 160~$\rm \mu m$
bands, which provides an approximate estimate for the intensity of the
diffuse emission in this direction. Using an average scaling factor of
1.27, the 100~$\rm \mu m$ flux at $l=30^\circ$ is estimated to be
1620~$\:{\rm MJy\:sr^{-1}}$. Our target IRDCs C, F, G are at
$l=28.3^\circ, 34.4^\circ, 34.8^\circ$, so we will use the results for
$l=30^\circ$. BT09 estimated that for IRDCs C, F, G the fraction of
diffuse emission that is in the foreground of the cloud is 0.266,
0.193, 0.14, respectively. Adopting a value of 0.2, so that 80\% of
the observed intensity is the value of the background intensity behind
the IRDCs, we have $I_{\nu,0}= 550,1290,1590\:{\rm MJy\:sr^{-1}}$.

\begin{figure*}[!tb]
\begin{center}
\includegraphics[width=6.5in,angle=0]{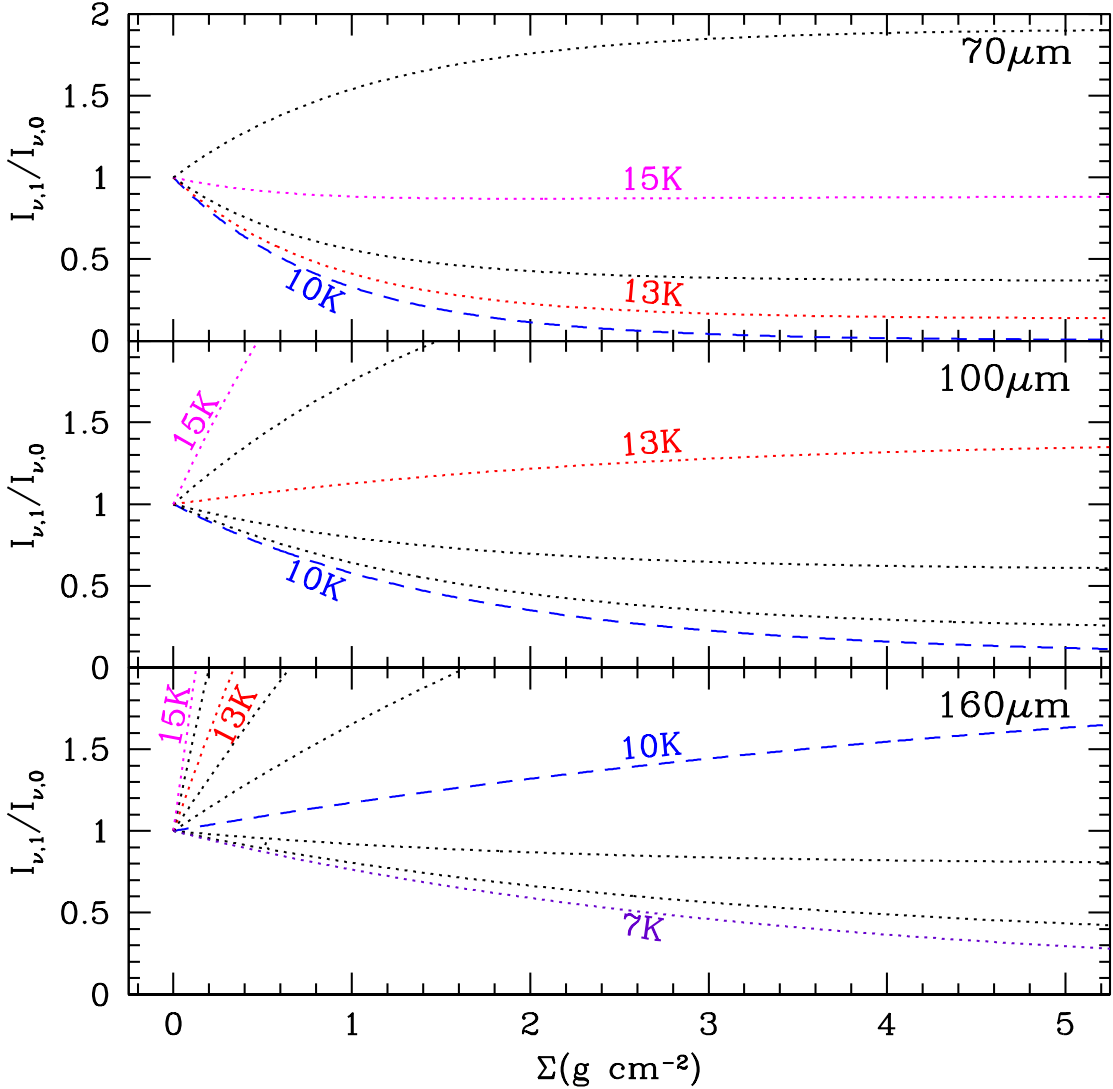} 
\end{center}
\caption{Dependence of $I_{\nu,1}/I_{\nu,0}$ with $\Sigma$ in the {\it
    Herschel}-PACS 70, 100, \& 160~${\rm \mu m}$ wavebands for IRDCs
  near $l=30^\circ$ and for various dust temperatures (the dotted lines
  show 1K increments between the labeled temperatures). OH94
  moderately coagulated thin ice mantle dust opacities have been
  adopted. Note the cloud needs to be cold ($\lesssim 15, 12, 9$~K) to
  appear dark at $70, 100, 160\:{\rm \mu m}$, respectively.
\label{fig:Iprofile}}
\end{figure*}

With the above fiducial values of $I_{\nu,0}$,
Figure~\ref{fig:Iprofile} shows $I_{\nu,1}/I_{\nu,0}$ as a function of
mass surface density, $\Sigma = \tau_\nu/\kappa_\nu$, for
different dust temperatures, $T_d$. We have evaluated
$I_{\nu,1}/I_{\nu,0}$ for the fluxes that would be received in the 70,
100, 160~$\rm \mu m$, i.e. the {\it Herschel}-PACS wavebands,
integrating over the filter response function, the Li \& Draine (2001)
spectrum of the diffuse Galactic background emission, and over the
OH94 opacity function (we find mean opacities of $\kappa_{\rm 70\mu
  m}=1.14\:{\rm cm^2\:g^{-1}}$, $\kappa_{\rm 100\mu m}=0.603\:{\rm
  cm^2\:g^{-1}}$ and $\kappa_{\rm 160\mu m}=0.290\:{\rm cm^2\:g^{-1}}$
for the OH94 moderately coagulated thin ice mantle dust model).

We examined images of the IRDCs in the {\it Herschel} data
archive. IRDC C has been imaged at 70, 100 \& 160~$\rm \mu m$. At
70~$\rm \mu m$, C1 appears globally dark (i.e. relative to the
low-intensity diffuse emission beyond the BT09 ellipse, not just
locally dark with respect to its immediate surroundings). At 100~$\rm
\mu m$ it has a similar, perhaps slightly lower, intensity as the
faintest parts of the IRDC surroundings, while at 160~$\rm \mu m$ it
appears moderately brighter (although is still locally
dark). Comparison with Fig.~\ref{fig:Iprofile} suggests a
(mass-weighted line of sight) temperature of $\simeq10$~K at C1. IRDC
F has images at 70 \& 160~$\rm \mu m$. F1 \& F2 are locally dark at
70~$\rm \mu m$. While F2 also appears to be globally dark, this does
not appear to be the case for F1 due to the relative proximity of
bright sources. At 160~$\rm \mu m$, the intensities towards both F1 \&
F2 are brighter than the diffuse surroundings. IRDC G has images at 70
\& 160~$\rm \mu m$. G2 appears globally dark at 70~$\rm \mu m$, but
brighter than the IRDC surroundings at 160~$\rm \mu m$.

Note that we expect the temperature in the $\rm N_2D^+$ cores to be
lower than the above constraints from $\sim 100\:{\rm \mu m}$
shadowing, which are mass-weighted averages along the line of sight
through the IRDC. Studies of lower-mass starless cores have seen
temperatures down to $\sim 6$~K in the very central regions
(e.g. Crapsi et al. 2007; Pagani et al. 2009). Given the above
constraints, we adopt a dust temperature of $T_d=10\pm3$~K for the
$\rm N_2D^+$ cores. As we will see, the densities are high enough that
we expect the gas and dust temperatures to be well-coupled, so we set
$T=T_d$.

We then evaluate $\Sigma_{\rm c,mm}$ via eq.~\ref{eq:Sigmamm}. Since
the uncertainties introduced by temperature are quite large and
asymmetric, we show upper and lower uncertainty bounds separately (we
have combined uncertainties in quadrature for the upper and lower
sides separately). The values of $\Sigma_{\rm c,mm}$ are always
greater than $\Sigma_{\rm c,min}$. They are less than $\Sigma_{\rm
  c,max}$, except in the case of C1-S, where it is about 5\% greater
than the BTK $\Sigma_{\rm c,max}$ value. These values still agree
given the uncertainties. 

We then proceed to derive masses ($M_{\rm c,mm}$) and densities
($n_{\rm H,c,mm}$) based on $\Sigma_{\rm mm}$. We will tend to regard
the core masses derived from mm dust emission as the most accurate
measure, rather than $M_{\rm c,max}$ or $M_{\rm c,min}$. This is
because the interferometric observations of the mm continuum emission
filter out contributions from structures $\gtrsim9$\arcsec, which are
the scales expected of the surrounding clump. The cores themselves
have diameters ranging from 3.7\arcsec\ to 7.3\arcsec. Note, that
$M_{\rm c,mm}$ still has uncertainties of about a factor of 2, due to
assumed temperature ($\pm3$~K), opacity (30\%) and distance (20\%)
uncertainties.


\begin{deluxetable}{cccccccc}
\tabletypesize{\footnotesize}
\tablecolumns{8}
\tablewidth{0pt}
\tablecaption{Identified \nndp Cores}
\tablehead{
\colhead{Core property (\% error)} & \colhead{C1-N} & \colhead{C1-S} & \colhead{F1} & \colhead{F2} & \colhead{G2-N} & \colhead{G2-S} & \colhead{Average}
}
\startdata
$l$ ($^\circ$) & 28.32503 & 28.32190 & 34.41923 & 34.43521 & 34.78097 & 34.77838 & ...\\
$b$ ($^\circ$) & 0.06724 & 0.06745 & 0.24598 & 0.24149 & -0.56808 & -0.56829 & ...\\
$\theta_c$ (\arcsec) & 3.38 & 3.61 & 2.49 & 1.87 & 3.67 & 2.31 & ...\\
$e$ & 0.771 & 0.214 & 0.794 & 0.666 & 0.706 & 0.583 & ...\\
P.A. ($^\circ$) & -40 & 77 & 115 & 111 & 135 & 66 & ...\\
\hline
$d$ (kpc) (20\%) & 5.0 & 5.0 & 3.7 & 3.7 & 2.9 & 2.9 & ...\\
 & &  & (1.56)\tablenotemark{b} & (1.56)\tablenotemark{b} & & &\\
\hline
$R_{\rm c}$ (0.01~pc) (20\%) & 8.18 & 8.75 & 4.46 & 3.35 & 5.16 & 3.25 & ...\\
 & &  & (1.88) & (1.41) & & &\\
\hline
\hline
$V_{\rm LSR,N_2D^+}$ ${\rm (km\:s^{-1})}$ & 81.18$\pm$0.03 & 79.40$\pm$0.01 & 56.12$\pm$0.01 & 57.66$\pm$0.04 & 41.45$\pm$0.01 & 41.80$\pm$0.01 & ... \\
$\sigma_{\rm N_2D^+,obs}$ ${\rm (km\:s^{-1})}$ & 0.367$\pm$0.032 & 0.365$\pm$0.016 & 0.172$\pm$0.011 & 0.376$\pm$0.047 & 0.288$\pm$0.013 & 0.240$\pm$0.015 & ... \\
$\sigma_{\rm N_2D^+,nt}$ ${\rm (km\:s^{-1})}$ & 0.363$\pm$0.032 & 0.361$\pm$0.016 & 0.164$\pm$0.012 & 0.372$\pm$0.047 & 0.283$\pm$0.013 & 0.234$\pm$0.015 & ... \\
$\sigma_{\rm N_2D^+}$ ${\rm (km\:s^{-1})}$ & 0.409$\pm$0.031 & 0.407$\pm$0.019 & 0.250$\pm$0.021 & 0.417$\pm$0.044 & 0.340$\pm$0.018 & 0.300$\pm$0.020 & ...\\
$V_{\rm LSR,DCO^+}$ ${\rm (km\:s^{-1})}$ & 81.18$\pm$0.02 & 79.59$\pm$0.02 & 56.26$\pm$0.01 & 57.53$\pm$0.02 & 41.52$\pm$0.01 & 41.91$\pm$0.02 & ...\\
$\sigma_{\rm DCO^+,obs}$ ${\rm (km\:s^{-1})}$ & 0.452$\pm$0.021 & 0.422$\pm$0.019 & 0.233$\pm$0.004 & 0.513$\pm$0.017 & 0.285$\pm$0.012 & 0.266$\pm$0.016 & ...\\
\hline
\hline
$\Sigma_{\rm cl}$ ${\rm (g\:cm^{-2})}$ (30\%) & 0.340 & 0.289 & 0.217 & 0.324 & 0.214 & 0.194 & ...\\
 & [0.525]\tablenotemark{a} & [0.442]\tablenotemark{a} & & & & &\\
\hline
$\Sigma_{\rm c,max}$ ${\rm (g\:cm^{-2})}$ (30\%) & 0.369 & 0.326 & 0.248 & 0.355 & 0.225 & 0.212 & ...\\
 & [0.609] & [0.514] & & & & &\\
\hline
$M_{\rm c,max}$ $(M_\odot)$ (50\%) & 37.1 & 37.5 & 7.41 & 5.97 & 9.01 & 3.36 & ...\\
 & [61.3] & [59.2]  & (1.32) & (1.06) & & &\\
\hline
$n_{\rm H,c,max}$ $(10^5{\rm cm}^{-3})$ (36\%) & 4.68 & 3.87 & 5.78 & 11.0 & 4.53 & 6.78 & ...\\
 & [7.73] & [6.10] & (13.7) & (26.1) & & &\\
\hline
$\sigma_{\rm c,vir,max}$ ${\rm (km\:s^{-1})}$ & 0.739$\pm$0.22 & 0.711$\pm$0.21 & 0.441$\pm$0.13 & 0.462$\pm$0.14 & 0.462$\pm$0.14 & 0.352$\pm$0.10 & ...\\
 & [0.934$\pm$0.28] & [0.887$\pm$0.26] & (0.287$\pm$0.085) & (0.300$\pm$0.089) & & &\\
\hline
$\sigma_{\rm N_2D^+}/\sigma_{\rm c,vir,max}$  & 0.554$\pm$0.17 & 0.573$\pm$0.17 & 0.565$\pm$0.17 & 0.902$\pm$0.28 & 0.736$\pm$0.22 & 0.853$\pm$0.26 & 0.697$\pm$0.088\\
 & [0.438$\pm$0.13] & [0.459$\pm$0.14] & & & & & [0.659$\pm$0.085]\\
 &  &  & (0.871$\pm$0.27) & (1.39$\pm$0.44) & & & ([0.791$\pm$0.11])\\
\hline
$R_{\rm c,vir,max}$ (0.01~pc) & 7.74$\pm$2.6 & 8.44$\pm$2.9 & 4.33$\pm$1.5 & 3.18$\pm$1.1 & 4.81$\pm$1.6 & 3.08$\pm$1.0 & ...\\
 & [8.00$\pm$2.7] & [8.57$\pm$2.9] & (1.82$\pm$0.62) & (1.34$\pm$0.46) & & &\\
\hline
$R_c/R_{\rm c,vir,max}$  & 1.06$\pm$0.42 & 1.04$\pm$0.41 & 1.03$\pm$0.41 & 1.05$\pm$0.41 & 1.07$\pm$0.42 & 1.05$\pm$0.42 & 1.05$\pm$0.17\\
 & [1.02$\pm$0.40] & [1.02$\pm$0.40] & & & & & [1.04$\pm$0.17]\\
\hline
\hline
$\Sigma_{\rm c,min}$ ${\rm (g\:cm^{-2})}$ (30\%) & 0.029 & 0.037 & 0.031 & 0.031 & 0.011 & 0.018 & ...\\
 & [0.084] & [0.072] & & & & & \\
\hline
$M_{\rm c,min}$ $(M_\odot)$ (50\%) & 2.92 & 4.26 & 0.926 & 0.522 & 0.44 & 0.286 & ...\\
 & [8.45] & [8.29] & (0.165) & (0.0927) & & &\\
\hline
$n_{\rm H,c,min}$ $(10^5{\rm cm}^{-3})$ (36\%) & 0.368 & 0.439 & 0.722 & 0.962 & 0.221 & 0.575 & ...\\
 & [1.07] & [0.854] & (1.71) & (2.28) & & &\\
\hline
$\sigma_{\rm c,vir,min}$ ${\rm (km\:s^{-1})}$ & 0.391$\pm$0.12 & 0.413$\pm$0.12  & 0.262$\pm$0.078 & 0.251$\pm$0.074 & 0.217$\pm$0.064 & 0.190$\pm$0.056 & ...\\
 & [0.569$\pm$0.17] & [0.542$\pm$0.16] & (0.170$\pm$0.050) & (0.163$\pm$0.048) & & &\\
\hline
$\sigma_{\rm N_2D^+}/\sigma_{\rm c,vir,min}$  & 1.05$\pm$0.32 & 0.986$\pm$0.30 & 0.951$\pm$0.29 & 1.66$\pm$0.52 & 1.57$\pm$0.47 & 1.58$\pm$0.48 & 1.30$\pm$0.17\\
 & [0.719$\pm$0.22] & [0.751$\pm$0.22] & & & & & [1.21$\pm$0.16]\\
 &  &  & (1.46$\pm$0.45) & (2.56$\pm$0.80) & & & ([1.44$\pm$0.20])\\
\hline
$R_{\rm c,vir,min}$ (0.01~pc) & 2.17$\pm$0.74 & 2.84$\pm$0.97 & 1.53$\pm$0.52 & 0.940$\pm$0.32 & 1.06$\pm$0.36 & 0.899$\pm$0.31 & ...\\
 & [2.97$\pm$1.0] & [3.21$\pm$1.1] & (0.645$\pm$0.22) & (0.396$\pm$0.13) & & &\\
\hline
$R_c/R_{\rm c,vir,min}$  & 3.77$\pm$1.5 & 3.08$\pm$1.2 & 2.91$\pm$1.1 & 3.56$\pm$1.4 & 4.86$\pm$1.9 & 3.61$\pm$1.4 & 3.63$\pm$0.59\\
 & [2.75$\pm$1.1] & [2.73$\pm$1.1] & & & & & [3.40$\pm$0.56]\\
\hline
\hline
$S_{\rm 1.34mm}$ (mJy) & 6.94$\pm0.72$ & 26.7$\pm0.77$ & 5.08$\pm0.63$ & 3.70$\pm0.30$ & 3.08$\pm0.48$ & 1.05$\pm0.31$ & ...\\
$S_{\rm 1.34mm}/\Omega$ (MJy/sr) & 8.25$\pm0.86$ & 27.7$\pm0.80$ & 11.1$\pm1.4$ & 14.4$\pm1.2$ & 3.10$\pm0.48$ & 2.66$\pm0.79$ & ...\\
\hline
$\Sigma_{\rm c,mm}$ ${\rm (g\:cm^{-2})}$ & $0.161^{0.321}_{0.0938}$ & $0.542^{1.08}_{0.322}$ & $0.218^{0.434}_{0.125}$ & $0.282^{0.560}_{0.165}$ & $0.0605^{0.121}_{0.0342}$ & $0.0521^{0.106}_{0.0260}$ & ...\\
\hline
$M_{\rm c,mm}$ $(M_\odot)$ & $16.2^{33.6}_{6.83}$ & $62.5^{129}_{26.8}$ & $6.51^{13.5}_{2.71}$ & $4.74^{9.80}_{2.01}$ & $2.42^{5.03}_{0.992}$ & $0.826^{1.74}_{0.296}$ & ...\\
 &  &  & ($1.16^{2.40}_{0.482}$) & ($0.842^{1.74}_{0.358}$) & & &\\
\hline
$n_{\rm H,c,mm}$ $(10^5{\rm cm}^{-3})$ & $2.05^{4.12}_{1.10}$ & $6.43^{12.9}_{3.52}$ & $5.07^{10.2}_{2.69}$ & $8.74^{17.6}_{4.72}$ & $1.22^{2.46}_{0.635}$ & $1.67^{3.41}_{0.766}$ & ...\\
 &  &  & ($12.0^{24.2}_{6.39}$) & ($20.7^{41.7}_{11.2}$) & & &\\
\hline
$\sigma_{\rm c,vir,mm}$ ${\rm (km\:s^{-1})}$ & $0.601^{0.828}_{0.418}$ & $0.808^{1.11}_{0.563}$  & $0.427^{0.589}_{0.297}$ & $0.436^{0.601}_{0.304}$ & $0.333^{0.459}_{0.231}$ & $0.248^{0.343}_{0.171}$ & ...\\
 & [$0.670^{0.923}_{0.466}$] & [$0.899^{1.24}_{0.626}$] & ($0.277^{0.383}_{0.193}$) & ($0.283^{0.390}_{0.197}$) & & &\\
\hline
$\sigma_{\rm N_2D^+}/\sigma_{\rm c,vir,mm}$  & $0.681^{0.983}_{0.487}$ & $0.504^{0.725}_{0.364}$ & $0.584^{0.845}_{0.416}$ & $0.956^{1.39}_{0.675}$ & $1.02^{1.48}_{0.736}$ & $1.21^{1.77}_{0.865}$ & $0.826^{0.986}_{0.725}$\\
 & [$0.611^{0.882}_{0.437}$] & [$0.453^{0.652}_{0.327}$] & & & & & [$0.806^{0.964}_{0.707}$]\\
 &  &  & ($0.899^{1.30}_{0.641}$) & ($1.47^{2.13}_{1.04}$) & & & ([$0.944^{1.13}_{0.826}$])\\
\hline
$R_{\rm c,vir,mm}$ (0.01~pc) & $5.12^{8.10}_{3.22}$ & $10.9^{17.2}_{6.90}$ & $4.06^{6.42}_{2.55}$ & $2.83^{4.48}_{1.79}$ & $2.49^{3.95}_{1.56}$ & $1.53^{2.44}_{0.925}$ & ...\\
 & [$4.12^{6.52}_{2.59}$] & [$8.80^{13.9}_{5.58}$] & ($1.71^{2.71}_{1.07}$) & ($1.19^{1.89}_{0.754}$) & & &\\
\hline
$R_c/R_{\rm c,vir,mm}$  & $1.60^{2.59}_{0.929}$ & $0.804^{1.30}_{0.468}$ & $1.10^{1.79}_{0.638}$ & $1.18^{1.91}_{0.687}$ & $2.07^{3.38}_{1.20}$ & $2.12^{3.58}_{1.22}$ & $1.48^{1.89}_{1.21}$\\
 & [$1.99^{3.22}_{1.15}$] & [$0.994^{1.60}_{0.578}$] & & & & & [$1.58^{2.08}_{1.30}$]\\
\hline
\enddata
\tablenotetext{a}{Properties derived from BTK MIR extinction map for
C1-N \& S, including sample averages, are shown inside ``[...]''.}
\tablenotetext{b}{Properties derived assuming the 1.56~kpc distance estimate to IRDC F (Kurayama et al. 2011), including sample averages, are shown inside ``(...)''.}
\label{tab:2}
\end{deluxetable}


\begin{deluxetable}{ccccccc}
\tabletypesize{\footnotesize}
\tablecolumns{7}
\tablewidth{0pt}
\tablecaption{Dynamical Properties of \nndp Cores\tablenotemark{a}}
\tablehead{
\colhead{Core property (\% error)} & \colhead{C1-N} & \colhead{C1-S} & \colhead{F1} & \colhead{F2} & \colhead{G2-N} & \colhead{G2-S}
}
\startdata
$R_{\rm c}$ (0.01~pc) (20\%) & 8.18 & 8.75 & 4.46 & 3.35 & 5.16 & 3.25\\
$\sigma_{\rm N_2D^+}$ ${\rm (km\:s^{-1})}$ & 0.409$\pm$0.031 & 0.407$\pm$0.019 & 0.250$\pm$0.021 & 0.417$\pm$0.044 & 0.340$\pm$0.018 & 0.300$\pm$0.020\\
$\Sigma_{\rm cl}$ ${\rm (g\:cm^{-2})}$ (30\%) & 0.525 & 0.442 & 0.217 & 0.324 & 0.214 & 0.194\\
$M_{\rm c,mm}$ $(M_\odot)$ & $16.2^{33.6}_{6.83}$ & $62.5^{129}_{26.8}$ & $6.51^{13.5}_{2.71}$ & $4.74^{9.80}_{2.01}$ & $2.42^{5.03}_{0.992}$ & $0.826^{1.74}_{0.296}$\\
$\alpha_{\rm c}\equiv 5\sigma_{\rm N_2D^+}^2 R_{\rm c}/(G M_{\rm c,mm})$\tablenotemark{b}  & $0.981^{2.35}_{0.419}$ & $0.270^{0.634}_{0.119}$ & $0.496^{1.20}_{0.209}$ & $1.43^{3.41}_{0.588}$ & $2.86^{7.04}_{1.24}$ & $4.13^{11.6}_{1.75}$\\
$n_{\rm H,c,mm}$ $(10^5{\rm cm}^{-3})$ & $2.05^{4.12}_{1.10}$ & $6.43^{12.9}_{3.52}$ & $5.07^{10.2}_{2.69}$ & $8.74^{17.6}_{4.72}$ & $1.22^{2.46}_{0.635}$ & $1.67^{3.41}_{0.766}$\\
$t_{\rm c,ff}$ $(10^5{\rm yr})$\tablenotemark{c} & $0.962^{1.31}_{0.678}$ & $0.542^{0.734}_{0.383}$ & $0.611^{0.838}_{0.431}$ & $0.465^{0.633}_{0.328}$ & $1.25^{1.73}_{0.878}$ & $1.07^{1.57}_{0.745}$\\
$B_c$ $({\rm \mu G})$ ($m_A=1$) & $156^{223}_{112}$ & $275^{390}_{202}$ & $115^{164}_{83.0}$ & $330^{474}_{234}$ & $94.4^{134}_{67.9}$ & $92.0^{132}_{61.8}$\\
$R_{\rm c,vir,mm}$ (0.01~pc) & $4.12^{6.52}_{2.59}$ & $8.80^{13.9}_{5.58}$ & $4.06^{6.42}_{2.55}$ & $2.83^{4.48}_{1.79}$ & $2.49^{3.95}_{1.56}$ & $1.53^{2.44}_{0.925}$\\
$R_c/R_{\rm c,vir,mm}$  & $1.99^{3.22}_{1.15}$ & $0.994^{1.60}_{0.578}$ & $1.10^{1.79}_{0.638}$ & $1.18^{1.91}_{0.687}$ & $2.07^{3.38}_{1.20}$ & $2.12^{3.58}_{1.22}$\\
$\sigma_{\rm c,vir,mm}$ ${\rm (km\:s^{-1})}$ & $0.670^{0.923}_{0.466}$ & $0.899^{1.24}_{0.626}$  & $0.427^{0.589}_{0.297}$ & $0.436^{0.601}_{0.304}$ & $0.333^{0.459}_{0.231}$ & $0.248^{0.343}_{0.171}$\\
$\sigma_{\rm N_2D^+}/\sigma_{\rm c,vir,mm}$  & $0.611^{0.882}_{0.437}$ & $0.453^{0.652}_{0.327}$ & $0.584^{0.845}_{0.416}$ & $0.956^{1.39}_{0.675}$ & $1.02^{1.48}_{0.736}$ & $1.21^{1.77}_{0.865}$\\
$\phi_{\rm B,vir}$ & $10.4^{25.5}_{3.91}$ & $23.1^{55.0}_{8.76}$ & $11.8^{29.0}_{4.39}$ & $3.16^{7.98}_{1.17}$ & $2.64^{6.35}_{0.992}$ & $1.68^{4.12}_{0.613}$\\
$m_{\rm A,vir}$ & $0.405^{0.758}_{0.249}$ & $0.262^{0.448}_{0.167}$ & $0.379^{0.697}_{0.233}$ & $0.899^{\infty}_{0.474}$ & $1.06^{\infty}_{0.545}$ & $1.99^{\infty}_{0.729}$\\
$B_{\rm c,vir}$ $({\rm \mu G})$ & $385^{677}_{176}$ & $1050^{1790}_{533}$ & $304^{534}_{141}$ & $367^{734}_{0}$ & $89.3^{181}_{0}$ & $46.3^{129}_{0}$\\
$B_{\rm c,crit}$ $({\rm \mu G})$ & $219^{478}_{134}$ & $736^{1610}_{452}$ & $296^{646}_{180}$ & $382^{834}_{234}$ & $82.2^{180}_{49.7}$ & $70.7^{156}_{40.7}$\\
\enddata
\tablenotetext{a}{Based on the mm continuum core masses, the BTK extinction map for
C1-N \& S, and the far distance for IRDC F.}
\tablenotetext{b}{Virial parameter (Bertoldi \& McKee 1992).}
\tablenotetext{c}{Core free-fall time, $t_{\rm c,ff} = [3\pi/(32G\rho_c)]^{1/2} = 1.38 \times 10^5 (n_{\rm H,c,mm}/10^5\:{\rm cm^{-3}})^{-1/2}\:{\rm yr}$.}
\label{tab:3}
\end{deluxetable}

\subsection{Dynamical state of the cores}\label{S:compare}

For each estimate of core mass, i.e. $M_{\rm c,max}$, $M_{\rm c,min}$
and $M_{\rm c,mm}$, we calculate $\sigma_{\rm c,vir}$ from
equation~(\ref{eq:s2}). In each case, we first consider the properties
of the cores and their clump envelopes as derived from the BT12 MIREX
maps. We also show the results for C1-N and S when using the BTK
extinction map, and the results for F1 and F2 when adopting the near
distance of 1.56~kpc (see Table~\ref{tab:2}). In addition to the
uncertainties in $\sigma_{\rm c,vir}$ due to errors in $M_c$ and
$\Sigma_{\rm cl}$, we also allow for a range of magnetic field
strengths in the core, such that $0.5<m_A<2$, i.e. $7.3>\phi_B>1.7$
(see \S\ref{S:theory}). We then evaluate the ratio of the observed
velocity dispersion derived from the $\rm N_2D^+$(3-2) spectrum to the
prediction from virial equilibrium, $\sigma_{\rm N_2D^+}/\sigma_{\rm
  c,vir}$. We average these values for the sample of 6 cores (assuming
the uncertainties are uncorrelated, although in reality there are
likely to be correlated systematic uncertainties, e.g. affecting mass
determinations due to choice of $\kappa$ or $T_d$). We present three
averages: first with core properties derived from the BT12 maps,
second with the C1 cores evaluated using the BTK maps (and the
remaining cores with the BT12 maps), and third being the same as the
second case, but with the F cores evaluated at the near distance. We
regard the second case as being the most accurate. We repeat the above
analysis for $R_{\rm c,vir}$ (evaluated from eq.~\ref{eq:rcore}) and
the ratio $R_c/R_{\rm c,vir}$ (note the distance uncertainty does not
affect $R_c/R_{\rm c,vir}$).

Quoting results for the second case, for $M_{\rm c,max}$ the core
sample has $\sigma_{\rm N_2D^+}/\sigma_{\rm c,vir}\simeq
0.659\pm0.085$, i.e. moderately sub-virial, while $R_c/R_{\rm
  c,vir}\simeq 1.04\pm 0.17$. Cores would appear to be sub-virial if
their masses have been overestimated. For $M_{\rm c,min}$, the sample
has $\sigma_{\rm N_2D^+}/\sigma_{\rm c,vir}\simeq 1.21\pm 0.16$,
i.e. moderately super-virial, and $R_c/R_{\rm c,vir}\simeq
3.40\pm0.56$. Cores would appear to have sizes larger than their
predicted equilibrium sizes if their masses have been
underestimated. Finally, for $M_{\rm c,mm}$ the sample has
$\sigma_{\rm N_2D^+}/\sigma_{\rm c,vir}\simeq 0.806^{0.964}_{0.707}$
and $R_c/R_{\rm c,vir}\simeq 1.58^{2.08}_{1.30}$. Given the estimated
uncertainties, we conclude that the sample of cores, with masses
estimated from mm continuum emission, have properties broadly
consistent with the virial equilibrium predictions of the Turbulent
Core Model. There is tentative evidence that the cores have velocity
dispersions that are marginally sub-virial, but as we shall see this is
degenerate with the assumed magnetic field strength in the cores.  The
core sizes appear to be slightly larger than the equilibrium size. We
note that this result may be influenced by our, somewhat arbitrary,
choice of defining the core size via the $3\sigma$ $\rm N_2D^+$(3-2)
contour.


Focusing on the most massive core, C1-S, with $M_{\rm c,mm}\sim
62.5^{129}_{26.8}\:M_\odot$, we find that $\sigma_{\rm
  N_2D^+}/\sigma_{\rm c,vir,mm}\simeq 0.453^{0.652}_{0.327}$ and
$R_c/R_{\rm c,vir,mm}\simeq 0.994^{1.60}_{0.578}$. This may indicate
that the core is quite strongly sub-virial, and so should be
undergoing fairly rapid global collapse (its free-fall time is only
$\sim 50,000$ years - Table~\ref{tab:3}). Alternatively, the core
could be closer to virial equilibrium and supported by stronger
large-scale magnetic fields than are assumed in the fiducial Turbulent
Core Accretion model.

The assumed fiducial value of the Alfv\'en Mach number $m_A= \sqrt{3}
\sigma_c/v_A = \sqrt{3} \sigma_c / (B/\sqrt{4\pi\rho_c}) = 1$ implies
a mean background field strength in a core of \beq B_c =
\sqrt{12\pi\rho_c} \sigma_c m_A^{-1} = 297 m_A^{-1} \left(\frac{n_{\rm
    H,c}}{10^5\:{\rm cm^{-3}}}\right)^{1/2} \left(\frac{\sigma_c}{\rm
  km\:s^{-1}}\right)\:{\rm \mu G}.
\label{eq:Bc}
\eeq In
Table~\ref{tab:3} we summarize the dynamical properties of the cores
(based on the mm continuum core masses, the BTK extinction map for
C1-N \& S, and the far distance for IRDC F), including $B_c$ for
$m_A=1$, evaluated using eq.~\ref{eq:Bc} with $\sigma_c=\sigma_{\rm N_2D^+}$. These values are $\sim 100 - 300\:{\rm \mu G}$.

To assess what field strength would be required for the core to be in
virial equilibrium with $\sigma_{\rm N_2D^+}/\sigma_{\rm c,vir,mm}=1$,
note that $\sigma_{\rm c,vir,mm} \propto \phi_B^{-3/8}$ (this assumes
that the same value of $\phi_B$ applies in the clump envelope, which
influences $\phipb$, as in the core). Thus in the case of C1-S, where
$\sigma_{\rm N_2D^+}/\sigma_{\rm c,vir,mm}=0.453$ (i.e. with
$\phi_B=2.8$), virial equilibrium would require $(\phi_{\rm
  B,vir}/2.8)^{-3/8} = 0.453$, i.e. $\phi_{\rm B,vir}=23.1$,
corresponding to $m_{\rm A,vir}=0.262$ and $B_{\rm
  c,vir}=1050\:{\rm \mu G}$. Analogous results for all the cores are shown in
Table~\ref{tab:3}. For the cores that appear to be sub-virial in the
fiducial analysis, a moderately stronger background magnetic field (by
factors of a few) is required for virial equilibrium. Note that the
ratio of $R_c/R_{\rm c,vir}$ is only weakly affected by a change in
$\phi_B$, but stronger large-scale magnetic field support should lead
to increased flattening of the core along the direction parallel to
the field lines.

Could the masses of the observed cores be set by their magnetic field
strengths? For an ellipsoidal core of length $2Z_c$ along the symmetry
axis and radius $R_c$ normal to the axis, the magnetic critical mass,
i.e. the maximum mass that can be supported by a magnetic field, $B_{\rm c,crit}$, is
(Bertoldi \& McKee 1992) \beq M_{\rm c,B} = 1.62
\left(\frac{R_c}{Z_c}\right)^2 \left(\frac{B_{\rm c,crit}}{100\:{\rm
    \mu G}}\right)^3 \left(\frac{n_{\rm H,c}}{10^5\:{\rm
    cm^{-3}}}\right)^{-2}\:M_\odot.
\label{eq:MBcrit}
\eeq If we equate $M_{\rm c,B} = M_{\rm c,mm}$ for our cores (assuming
$R_c=Z_c$), we can estimate the values of $B_{\rm c,crit}$ that apply
to their particular masses and densities. We find that these values
are similar to $B_{\rm c,vir}$ for all six cores, even though there is
a range of required Alfv\'en Mach numbers, $m_{\rm A,vir}$ of a factor
of about eight. If a moderate degree of flattening is assumed so that
$R_c=2Z_c$, then the estimates of $B_{\rm c,crit}$ decrease by a
factor of $(Z_c/R_c)^{2/3}\rightarrow 0.63$ (at fixed $n_{\rm H,c}$).

Unfortunately, it is difficult to observationally determine magnetic
field strengths in IRDC cores and we do not have any direct
constraints for these particular sources. Measurement of Zeeman
splitting from molecules such as CN (Falgarone et al. 2008) require
relatively strong lines, whereas the observed emission is typically
quite weak. From 14 regions, with average density of $n_{\rm
  H}=9\times 10^5\:{\rm cm^{-3}}$, Falgarone et al. derived a median
value of the total field strength of $560\:{\rm \mu G}$. Such values
are similar to those needed for our $\rm N_2D^+$ cores to be in virial
equilibrium. Our most massive core requires a moderately higher value
(and at a slightly lower density), but since massive starless cores
(that will form massive stars) are rare objects, it is quite possible
they require relatively unusual (stronger) magnetic field
strengths. The Falgarone et al. magnetic field strength measurements
form part of a set that were used by Crutcher et al. (2010) to
estimate a median field strength versus density relation
\beq
B_{\rm med} \simeq 0.22 \left(\frac{n_{\rm H}}{10^5\:{\rm cm^{-3}}}\right)^{0.65}\:{\rm mG}\:\:\:(n_{\rm H}>300\:{\rm cm^{-3}}),
\label{eq:crutcher}
\eeq 
with uniform distribution of values up to $2 B_{\rm med}$.  This
predicts a median field strength about 1.6 times stronger than the
Falgarone et al. values. Applying this relation to the density of C1-S
yields $B_{\rm med} = 730\:{\rm \mu G}$, close (within $\sim 40\%$) to
the value needed for virial equilibrium. Formally, about 30\% of the
Crutcher et al. uniform distribution of field strengths at this
density would be strong enough to support the core, but the
uncertainties associated with this estimate, which is also model
dependent, are large.

Magnetic field morphology can be measured via dust continuum
polarization, but there are large uncertainties in estimating field
strengths via the Chandrasekhar-Fermi method. Relatively order
magnetic field morphologies, and thus relatively strong fields
strengths, have been claimed to be present around some massive
protostars (e.g. Girart et al. 2009), which is additional indirect
evidence that dynamically important fields are also present at the
earlier, starless core stage.

Another effect to consider is the possibility that, if magnetic
fields are dynamically important, the true velocity dispersions of the
cores are underestimated by those observed in ionized species,
such as $\rm N_2D^+$ (Houde et al. 2000; Falceta-Gon\c{c}alves et
al. 2010; Tilley \& Balsara 2010). Observations of neutral species in
the core are needed, but many species are likely to have frozen-out
onto dust grain ice mantles (as discussed above, we did not detect
DCN(3-2) or $^{13}$CS(5-4) towards the cores). If co-spatial neutral
species emission is observed to have a larger velocity dispersion this
would be evidence in support of dynamically important B-fields.


The cores discussed in this paper are overdense structures, most
likely driven to their current structure by their self-gravity. Thus
there is a strong possibility they are still in a state of
contraction, which would deviate from that of perfect virial
equilibrium. Signatures of infall, e.g. as traced via the asymmetric
profiles of optically thick lines, need to be searched for. Models of
collapsing cores, even if initial regulated by magnetic fields, do
predict magnetically supercritical central regions (e.g., Ciolek and
Mouschovias 1994).

\subsection{Comparison to some previous studies}

Csengeri et al. (2011) studied the dynamics of five massive
protostellar cores (i.e. those that are already forming a protostar)
in Cygnus-X, concluding that the velocity dispersions were smaller
than the predicted MT03 values for a virialized core. They measured
velocity dispersions from observations of $\rm H^{13}CO^+$(1-0)
($\sim4\arcsec$ angular resolution; 0.13~$\rm km\:s^{-1}$ velocity
resolution) and $\rm N_2H^+$(1-0) (29\arcsec\ angular resolution;
0.13~$\rm km\:s^{-1}$ velocity resolution). They assumed a single
value of $\Sigma_{\rm cl}=1.7\:{\rm g\:cm^{-2}}$ based on mm continuum
emission. The main differences compared to our present study include:
(1) they studied cores that are already forming stars; (2) they did
not use deuterated species to trace the core gas; (3) they measured
clump properties via mm continuum emission, rather than MIR extinction
mapping.

Pillai et al. (2011) studied the dynamics of cold cores in the
vicinity of ultracompact HII regions, also concluding they were
strongly sub-virial. They used $\rm NH_2D$ emission
($\sim4.5$\arcsec\ angular resolution; 0.27~$\rm km\:s^{-1}$ velocity
resolution) to measure velocity dispersion and 3.5~mm continuum
emission to estimate mass. The main differences compared to our
present study include: (1) Many of their cores are unresolved, with
estimated deconvolved sizes that are much smaller than the angular
resolution of their observations - yet all mm emission, i.e. mass, is
assumed to originate from inside these effective radii. On the other
hand, all of our cores are resolved. (2) No allowance was made for
surface pressure terms in the consideration of virial equilibrium.

\section{Conclusions}

We used ALMA Cycle 0 observations to search for $\rm N_2D^+$(3-2)
emission from four IRDCs, selected to be dark at wavelengths as long
as 70~$\rm \mu m$. Strong detections were made in all cases, leading
to identification of 6 $\rm N_2D^+$ cores. We assessed their
properties, including their surrounding clump envelopes, via MIR
extinction maps. We also measured core masses via the detected 1.34~mm
dust continuum emission. We regard this mass estimate as the most
accurate for the cores (even though it still shows factor of 2
uncertainties), because of the difficulty of separating out core from
clump material in the MIREX-derived mass surface density maps. The
MIREX maps do, however, allow us to estimate the clump envelope
properties, which are needed for the surface pressure term in the
virial equation. 

We assessed the dynamical state of the cores by comparing to the
Turbulent Core Model of MT03. For the sample of 6 cores the ratio of
the observed to the predicted (virial equilibrium) velocity dispersion
is $0.81^{0.96}_{0.71}$, while the ratio of the observed to predicted
size is $1.58^{2.08}_{1.30}$. Note, these error intervals assume
random errors amongst the six cores, but if there are correlated
systematic errors, e.g. in mass estimates, then the true error range
could be larger. Thus, given the uncertainties, the cores appear to be
quite close to the predictions of the model, which assumes virial and
pressure equilibrium and invokes magnetic fields such that the
Alfv\'en Mach number is $m_A=1$. There is tentative evidence that the
cores are slightly sub-virial compared to the fiducial model,
especially in the case of the more massive cores C1-S and C1-N, with
$\sim 63\:M_\odot$ and $\sim 16\:M_\odot$, respectively. However,
these cores could be close to virial equilibrium if they are threaded
by moderately stronger large-scale background magnetic fields with
strengths up to $\sim 1$~mG, implying that $m_A\simeq 0.3 - 0.4$.

To prevent fragmentation of the cores requires similar field
strengths, which we have evaluated by equating observed core mass with
the magnetic critical mass given its observed density.
This may indicate that magnetic fields are dynamically important in
setting the mass function of cores, as has been suggested by Kunz \&
Mouschovias (2009) (see also Bailey \& Basu 2013), especially preventing the fragmentation of the
massive cores that eventually form massive stars. Magnetic suppression
of fragmentation appears to be the most likely mechanism operating in
IRDCs, since the cold temperatures ($\sim 10$~K) indicate that strong
radiative heating from surrounding protostars to raise the thermal
Jeans mass (Krumholz \& McKee 2008) is not occurring. Magnetically
mediated massive star formation would not require a minimum mass
surface density threshold, such as the $\Sigma_{\rm cl}\simeq 1\:{\rm
  g\:cm^{-2}}$ limit proposed by Krumholz \& McKee. Indeed the clump
medium immediately surrounding C1-N and C1-S only has $\Sigma_{\rm
  cl}\simeq 0.5\:{\rm g\:cm^{-2}}$ and the clumps in the IRDC sample
of BT12 tend to have even lower values.

This pilot study with ALMA revealed two relatively massive starless
cores out of a total sample of six detected objects. Further
observations of other regions are needed to build a larger statistical
sample. Observations to constrain the magnetic field strengths in
these objects are also needed.

\acknowledgments We thank Gary Fuller, Jouni Kainulainen, Wanggi Lim,
Chris McKee, Andy Pon and Scott Schnee for helpful discussions, as well as the
comments of an anonymous referee. JCT acknowledges support from NSF
CAREER grant AST-0645412; NASA Astrophysics Theory and Fundamental
Physics grant ATP09-0094; NASA Astrophysics Data Analysis Program
ADAP10-0110. SK acknowledges support from an NRAO-SOS grant in support
of ALMA-Cycle 0 observations. This paper makes use of the following ALMA data: 
ADS/JAO.ALMA\#2011.0.00236.S. ALMA is a partnership of ESO (representing 
its member states), NSF (USA) and NINS (Japan), together with NRC 
(Canada) and NSC and ASIAA (Taiwan), in cooperation with the Republic of 
Chile. The Joint ALMA Observatory is operated by ESO, AUI/NRAO and NAOJ.
The National Radio Astronomy Observatory is a facility of the National 
Science Foundation operated under cooperative agreement by Associated 
Universities, Inc.



\end{document}